\documentstyle[ip]{article}

\font\twelvero=cmr12
\def\V{\cal V}
\def\NPB{{\em Nucl. Phys.} B}
\def\be{\begin{equation}}
\def\ee{\end{equation}}
\def\bea{\begin{eqnarray}}
\def\eea{\end{eqnarray}}

\begin{document}

\title{
\vspace{-5.5cm}
\begin{flushright}{\normalsize RU-99-8}\end{flushright}
\vskip 4.5cm
\twelvero
\bf {Mathematical aspects of chiral
gauge theories on the lattice\footnote{Invited talk at
``Trends in Mathematical Physics'', Oct. 14-17, University of
Tennessee, Knoxville, USA.}
}}

\author{\twelvero{Herbert Neuberger}}
\address{\twelvero{Department of Physics and Astronomy\\
Rutgers University,\\ Piscataway, NJ 08855-0849, USA \\
E-mail: neuberg@physics.rutgers.edu}}

.
\vskip1.4truein
\maketitle

\centerline{\bf Abstract}
\begin{minipage}{4.0truein}
                 \footnotesize
\leftskip .5in
\par
For two decades it was believed that chiral
symmetries cannot be realized in lattice field
theory but this has changed now. 
Highlights of these new developments will be
presented with emphasis on the mathematical
structure of the so called ``overlap''.         
\end{minipage}
         \vskip 2em \par\goodbreak

\section {Introduction}

Up to energies $E$, tested in present day's most powerful
accelerators, one finds that Nature can be accurately
described by a ``standard model'', 
a chiral gauge theory with a small
non-vanishing short-distance cutoff ${1\over\Lambda}$ 
of unknown form but known
magnitude. 
Accurate predictions are possible, 
despite ignorance about short distances,  
because the theory is renormalizable. 
To
accuracy ${E\over \Lambda}\approx 10^{-12}$ the 
influence of
the short distance structure can be entirely absorbed
in a few measurable parameters the 
chiral gauge theory depends
on. Future experiments will increase $E$,
and possibly change the standard model.
Very likely, the model will just evolve
into another chiral
gauge theory, maybe supersymmetric. 

Renormalizability is rigorously established 
order by order in ``perturbation theory'' (a certain
asymptotic expansions of the theory), but
is more difficult to prove for the
theory as a whole.
Mainly because the theory is chiral, 
renormalizability outside perturbation theory
is problematic. 

Although not established with
full mathematical rigor, the accepted view is that
any theory is 
renormalizable outside perturbation theory 
if one can replace its true short distance structure
by a discretization obeying certain general 
restrictions. 
When the discretization preserves enough of
the symmetries one extracts 
physical information by going
to a ``continuum limit'' where short distance
details of the construction become irrelevant. 
It is relatively easy to discretize ``vectorial''
gauge theories, but chiral gauge theories pose new
difficulties.
There, perturbative
renormalizability hinges on a 
delicate cancelation (of ``anomalies'') and it
is difficult to incorporate this
cancelation at a fundamental level 
in lattice formulations. Until quite recently,
because of several ``no go theorems'', it was suspected 
by some 
that chiral gauge theories may not be renormalizable
outside perturbation theory. This belief has 
lost many subscribers by now.

A less
severe, but related problem occurs even in the
context of vectorial gauge theories where
almost exact global chiral symmetries (i.e.
``not gauged'') are known to hold. Traditional
lattice regularizations require an unnatural
and cumbersome fine tuning to achieve global chiral
symmetries in the continuum limit. Until recently it was
thought that one cannot preserve 
exact global chiral symmetries 
before the continuum limit is taken. 

One does not need to study the renormalizability 
of the complete gauge-matter
system in order to deal with the problem of chirality.
Unlike in renormalization theory, the specific issues
one needs to resolve when quantizing chiral fermions
are almost independent of the (even) dimensionality 
of the model. The problem of 
quantizing chiral fermions in a
fixed gauge field background can be formulated 
in any even Euclidean dimension. 

Renormalizability of the theory
as a whole is an issue that can 
be separated from the problem of 
chirality.  
The quantization of gauge-matter
systems can be split into two steps: In the first
only the fermions are quantized, while the
gauge field background is kept fixed. In the second, 
also the gauge fields are quantized. We only
need to worry about the
first step. This step, in isolation, has trivial
renormalization - and can be analyzed in any (even)
dimension. The second step typically requires Euclidean
dimension less than or equal to four. 

If one
so chooses one can deal with the quantization
of the entire system as a whole. But, in my
opinion, this would be a bad idea. A crucial
simplification is afforded by fermionic matter
entering the action only 
bilinearly and this simplification is
only exploited fully in the two step approach. 

\section {The lattice and the overlap}

Our mathematical problem can be described as follows.
Start from a $2d$ dimensional compact
Riemannian manifold $M$. It 
is the base space of a principal
fiber bundle with structure group ${\cal G}$. We also
have an associated spin bundle and the Dirac operator
$D$ acting on its sections. The 
spin bundle can be decomposed
into its chiral components and 
there are two Weyl ($W_{L,R}$) operators
connecting corresponding sections. We shall simplify
our discussion to flat Riemannian manifolds with the
topology of a $2d$-dimensional torus, $T^{2d}$. 
Also, to be specific, we shall pick ${\cal G}=SU(n)$.
We replace the torus by a hypercubic finite lattice with
sites $x$. Connections on the fiber bundle
are replaced by discrete collections of elementary
parallel transporters $U(x,y)$.
Parallel transport along a sequence of arcs $(x,y)$ in
the graph associated with the lattice is implemented
by an ordered product of $U(x,y)$ along the path.
The individual parallel transporters are $SU(n)$ matrices. 
A collection of ``link variables'' $U$ makes up the
``gauge background''. At each site of the lattice
we can perform an $SU(n)$ rotation on the local frame,
inducing a ``gauge transformation'' on the connection.
A collection of connections 
related by all possible gauge transformations
constitutes a ``gauge orbit''. 
In the continuum limit, physical
degrees of freedom are associated only 
with orbits, not with
individual connections. 

In the continuum, physical results require an integration
over all orbits with certain weight functions. The
integrals are written as integrals over connections
but the integrands are picked gauge invariant, i.e. 
dependent only on the orbit. This integration
is carried out as part of step 
two in the construction program.
Step one is required to provide two kinds of quantities,
both determined by the gauge background: The first
one is a function and plays the role 
of the determinant of the Weyl
operator. The determinant is expected to be gauge
invariant. The other class of quantities 
contains the matrix
elements of the inverse of the Weyl operator 
evaluated in a standard basis. These are the  
``fermionic Green's function'' or ``propagators''. 
The Green's functions
must transform by conjugation under gauge transformations.
One cannot
pick arbitrarily independent 
definitions for the determinant
and the fermion Green's functions: 
Variations of the determinant
under small deformations of the 
gauge background have to obey
natural expressions in terms of the 
fermion Green's functions. 

On the lattice
we would expect to replace the continuum operators by 
finite matrices with a standard definition of
finite dimensional inner products replacing
the continuum inner product in the Hilbert space
of sections. 
However, since the Weyl operators
can have a nonzero analytical index which 
depends on the gauge field, one cannot
decide in advance on the shape of the matrices.
This observation, although quite trivial, 
is relatively new and
was crucial to the progress I am reporting on.

The disconnected space of
connections we are familiar with in
the continuum is replaced by a product
of $SU(n)$ group manifolds, one for each link. 
Thus, the ``parameter
space'' of our matrices is now
connected and cannot admit a smooth definition
of ``shape'' except when it 
is constant. This difficulty faced by lattice
regularizations is well
understood since the work of Phillips 
and Stone on ``lattice
topology''; one
needs to excise from the set of all lattice
gauge backgrounds some subset, preferably,
but not necessarily, of zero measure. This leaves
behind a more complicated parameter space, cut up
in disconnected pieces. On each piece the shape
of the matrix representing the Weyl
operators is fixed. The  
shape is defined by an integer
determining the difference between the number
of rows and columns, $2k$. 
The sum of the two, $2K$ ($K\ge|k|$), can
and is kept fixed, determined 
solely by the lattice and $n$. Each
matrix is paired with a complementary one,
with opposite sign of $k$. The complementary
matrix is associated with Weyl fermions
in the conjugate representation, or equivalently,
with fermions of opposite handedness. 
The two matrices can be assembled together into
a $2K\times 2K$ matrix (each side of this matrix
has a length equal to the sum of the numbers
of rows and of columns of either Weyl matrix) 
representing Dirac fermions. The bigger matrix has
a fixed shape, independently of the gauge
background. The main reason why 
vectorial gauge theories are
so much easier to define is that the Dirac operator
can be replaced by a matrix of fixed shape. However,
to preserve exact masslessness one needs to preserve
exact decoupling between left and right components. Since
the two Weyl blocks cannot have a 
fixed shape any discretization
preserving exact masslessness of 
Dirac fermions on the lattice
has a Dirac matrix which is not 
analytic in the link parallel
transporters.\footnote{This has some interesting
consequences: It goes without saying that 
the Dirac matrix must change by conjugation
under gauge transformations. In turn, 
this means that the Dirac matrix connects
two distinct sites by a sum over path ordered products
of elementary link matrices. All this goes through also for
a structure group given by $U(1)$. In this case, one can 
pick as an example all parallel 
transporters in a fixed direction
as equal to each other, given 
by $e^{ip_\mu},~\mu=1,2,...,2d$. Our
discussion implies that the resulting reduced 
Dirac matrix cannot be analytic on the
torus spanned by the $p_\mu$'s. 
This implies that the Dirac
matrix cannot be sparse in site space.}

The ``overlap'' provides a construction of
matrices representing Weyl operators. 
One starts with the easier problem of defining on
the lattice a certain massive Dirac
operator. One picks 
an appropriate Hermitian 
operator $H_W$ which transforms covariantly 
under lattice gauge transformations and represents
massive Dirac fermions. 
There are no difficulties because chiral symmetries
are absent for massive fermions. There
is a large amount of latitude in choosing the structure
of $H_W$: the single requirement is that $H_W$ be a 
lattice approximation to a massive Dirac
operator with a {\it negative} and large mass
term.\footnote{On the lattice the actual 
distinction between a positive and negative mass is
that the massive lattice Dirac operator with positive
mass can be smoothly deformed to infinite positive mass
without ever becoming singular. The negative mass operator 
does not admit such deformations.} One excludes
all lattice gauge field backgrounds for which
$H_W$ has a nonzero kernel. (On the lattice, if one
chose a negative mass term, this kernel will end
up being non-empty for many backgrounds.) This exclusion
is gauge invariant; in other words gauge orbits
are excluded in their entirety. 
Over the remaining space of gauge backgrounds 
one unambiguously splits the finite vector
space $\V$, on which $H_W$ acts, into $\V_+\oplus
\V_-$ where $\V_\pm$ are the images of $\V$ under
${{1\pm \varepsilon (H_W )} \over 2}$.
$\varepsilon(x)$ is the sign function. $\V$ also
has a canonical split 
into $\V^\prime_+\oplus
\V^\prime_-$ induced 
by $\gamma_{2d+1}\equiv \epsilon^\prime$. The
chirality operator $\epsilon^\prime$ 
also splits the spin bundle 
in the continuous case. There, it
splits the Dirac
operator into Weyl operators.
Here, the lattice Weyl operators are represented by
``overlap'' matrices between the
corresponding subspaces $\V^\prime_+$ and $\V_+$ for
one handedness and $\V^\prime_-$ and $\V_-$ for the
other. The massless Dirac operator is realized on the
lattice as a combination of the two, and is
given by 
$D_o = {{1+\gamma_{2d+1}\varepsilon(H_W )}\over 2 }$. 
Alternatively, one deals with its hermitian version,
$H_o = \gamma_{2d+1} D_o = {{\gamma_{2d+1}+\varepsilon(H_W )}\over 2 }
\equiv {{\epsilon^\prime +\epsilon}\over 2}$.

\section {Vectorial gauge theory and topology}

The continuum massless Dirac operator maps elements
of $\V^\prime_\pm$ (left/right components)
into elements of $\V^0_\mp$. For
this reason, if there are several copies (flavors)
of massless
Dirac operators their left and right components 
can be complex-rotated 
into each other independently without changing
the determinants. A flavor
independent mass term
would prohibit independent rotations, and only
a vectorial symmetry, with both rotations equal to each
other, would be allowed. The technical  
difference between the massive and the massless
Dirac operators is that the latter 
anticommutes with $\epsilon^\prime$. 

Since $\epsilon$ and $\epsilon^\prime$ are hermitian
and square to one, the 
operator $V=\epsilon^\prime \epsilon$
is unitary and obeys $ \epsilon^\prime V\epsilon^\prime
=V^\dagger$. Actually, if we only know that we have
a unitary operator $V$ with the above property, we
can define $\epsilon$ so that it be 
hermitian and square to one. Unlike the continuum
Dirac operator, the lattice
overlap Dirac operator $D_o \equiv {{1+V}\over 2}$,
does not anticommute with $\epsilon^\prime$. 

All the important
consequences of chiral symmetry are expressed
in terms of identities between products
of matrix elements of the inverses of the Dirac
operator (propagators). Now, $D_o^{-1} ={2\over {1+V}}=
{{1-V}\over{1+V}} +1$. The first term,
${{1-V}\over{1+V}}$, is easily seen to anticommute
with $\epsilon^\prime$. Thus, the single violations
of the chiral identities will 
occur for identities involving 
diagonal elements of the inverses. There are
many chiral identities, and most 
do not involve diagonal elements
of the inverses. Actually, 
the violations are immaterial for the continuum limit;
in ordinary field theory one 
always expects singularities
to arise when operators are multiplied at the same
point.\footnote{The local operators are more 
accurately described as operator
valued distributions.}
In the process of constructing the 
algebra of local operators 
in the continuum 
these singularities need to be redefined anyhow.
Thus, for anything that really matters, chiral
symmetry is preserved even with $D_o$. 

That chiral symmetry would
work on the lattice by allowing some local violations in
the propagators was observed a long time
ago by Ginsparg and Wilson 
who formalized the requirement
in a certain relation; however they did not
produce an explicit example for 
an acceptable lattice Dirac operator
satisfying their relation in 
arbitrary gauge backgrounds.
They showed that a lattice action will essentially have
exact global chiral symmetry if 
the fermion matrix, $D_{GW}$,
obeyed the 
requirements $\{\gamma_5 , D_{GW}^{-1} - R\} =0$
($\{..,..\}$ denotes the anticommutator), 
with $R=R^\dagger,~[\gamma_5, R]=0$
and $\gamma_5 D_{GW}$ hermitian, 
while $R$ is a ``local'' operator,
strongly diagonally dominated in site space. 
Defining $D^{-1}_c = D_{GW}^{-1} -R$
and the unitary matrix $V={{1-D_c}\over{1+D_c}}$ 
we find that the
most general solution of the 
Ginsparg-Wilson requirements is
$D_{GW}^{-1} = D_o^{-1} +R-1$, 
where $D_o$ is an overlap Dirac
operator. Apparently, there is no 
advantage in picking an $R$ different from
the unit matrix, and one often restricts $D_{GW}$ 
to $D_o$.
The crux of the matter is 
to find a unitary matrix $V$
for which $\epsilon =\epsilon^\prime V$ 
is hermitian such that
the associated $D_o$ be a faithful 
approximation to the massless
continuum Dirac operator. 
Equivalently, one can look for 
a hermitian
$\epsilon$ which squares to unity
so that ${{1+\epsilon^\prime \epsilon}\over 2}$
is an acceptable approximation to the 
massless continuum Dirac operator.

When the gauge background is a connection on
a nontrivial bundle over the torus, the continuum Dirac
operator has exact zero modes because its Weyl
components have a nontrivial analytic index. On
the lattice the associated 
topological integer is ${1\over 2}
tr\epsilon$. 
One can prove that the associated zero
modes exist as follows. Consider two
orthonormal bases, 
$\{\psi^\prime_j |\epsilon^\prime
\psi^\prime_j=\epsilon^\prime_j \psi^\prime_j,
 j=1,2,...2K\}$ and
$\{\psi_j |\epsilon
\psi_j=\epsilon_j \psi_j,
 j=1,2,...2K\}$. Obviously,
the $\psi_j$ can be chosen as also 
eigenvectors of $H_W$. Exactly
$K$ of the $\epsilon^\prime_j$'s are unity. 
One has $(\psi^\prime_j, H_o \psi_i)
=(\psi^\prime_j ,\psi_i) 
{{\epsilon^\prime_j +\epsilon_i}\over 2}$ so the
rank of $D_o$($=\epsilon^\prime H_W$) is generically
equal to $2K-|tr\epsilon |$, as
expected. For generic backgrounds 
the rank will not change under small but arbitrary
deformations of the connection.

It is easy to write down 
operators $\epsilon ,V$ and their associated
$D_o$, which simply do not 
represent massless fermions
and also are insensitive to topology, 
but nevertheless 
obey  the Ginsparg-Wilson relation. 

\section{Propagators, the determinant bundle and anomalies}

With the help of the 
operators $\epsilon$ and $\epsilon^\prime$
one can decompose the total space $\V$ into a direct sum
of orthogonal subspaces 
in two ways: $\V=\V^\prime_+\oplus\V^\prime_-$
with $\epsilon^\prime=\pm 1$ on $\V^\prime_\pm$ or
$\V =\V_+ \oplus\V_-$ 
with $\epsilon=\pm 1$ on $\V_\pm$. The second
decomposition is well defined only for backgrounds
where $H_W$ is invertible. Let 
the associated orthogonal projectors be
$P^\prime_\pm , P_\pm$. Let us 
think about the projectors
as maps from $\V$ to the respective 
subspaces. So, if bases are chosen,
the projectors $P^\prime_\pm$ 
would be represented by rectangular matrices with
$2K$ columns and $K\pm k$ rows while the projectors $P_\pm$ 
would be represented 
by $K\times 2K$ matrices. The Weyl operator 
$W^{L}: \V^\prime_+ \to \V_+$ 
is given by $P_+ {P^\prime_+}^\dagger$ and 
$W^{R}: \V^\prime_- \to \V_-$ 
is given by $P_- {P^\prime_-}^\dagger$. 
Their ``inverses''
(propagators) are extended 
to $\V$ as $G^L= {P^\prime_+}^\dagger
{1\over{P_+ {P^\prime_+}^\dagger}} P_+$
and $G^R= {P^\prime_-}^\dagger
{1\over{P_- {P^\prime_-}^\dagger}} P_-$. 
These propagators
have ranks $K\pm k$.  

The subspaces $\V_\pm$ are
defined in a gauge invariant way: the subspaces do not
change along gauge orbits, so can be viewed as defined
over the space of allowed gauge connections modded out by
gauge transformations. Intuitively,
$W^L$ measures a ``distance'' between the spaces
$\V^\prime_+$ and $\V_+$. If $W^L$ is a unit matrix
(up to phases) the spaces coincide. 

The ``determinant'' has to be $\det W^L$. 
It can be nonzero only
for zero topology, since otherwise $W^L$ isn't
square. Although the map $W^L$ is
completely determined by the 
subspaces $\V^\prime_+$
and $\V_+$ (which are defined 
in a gauge invariant
way), since it connects two 
different spaces, there
is no natural numerical determinant one 
can associate to it.
This mirrors the situation in continuum. 
What we really end up with
is a definition of a line bundle 
over the space of admissible 
gauge orbits, not a function. The problematic
part is a determinant bundle given by the 
collection of all spaces 
$\V^{\wedge}_+ =\V_+ \wedge \V_+ 
\wedge \dots \V_+$, where the number
of factors is the dimension of $\V_+$. 
In physics, a choice
of basis in a $\V^{\wedge}_+$ is a ``ground state'' in a
``second quantized fermionic system'' 
with strictly bilinear
interactions given by $H_W$. 
It is crucial that $W^L$
is constructed from $H_W$ which is $C^\infty$ in
the gauge background, even before any gauge configurations
have been excised. We only have a line bundle, so  
the phase of the complex 
number $\det W^L$ is still undefined. 
The bundle base space is the collection 
of gauge orbits, so
gauge invariance is automatic but there may be no smooth 
sections. In other words, 
while $|\det W^L|=|det(P_+ P^\prime_+ )|$ 
is gauge invariant, 
it is not yet sure that $\det W^L$ can 
also be made gauge invariant
smoothly in the background gauge orbits. 
This smoothness is necessary
for meaningful quantization.

To make progress we need to go back a step and work over
the space of connections rather than gauge orbits. This
space is contractible in each topological sector. We 
have the same fibers as before. Now we pick a section. 
The section naturally produces a one form over base space.
This one form can be viewed as a 
connection in a $U(1)$ bundle.
In Physics it is known as Berry's connection. In our case
Berry's connection corresponds to a known function
of the gauge background in the continuum, namely the
difference between the ``covariant anomaly'' and
the ``consistent anomaly''. When this difference is 
nonzero, any definition of a
gauge invariant chiral determinant becomes untenable in the
continuum. On the lattice it is suspected that, if Berry's
connection turned out to vanish, the section it came from
could be made gauge invariant. 
Then we could take this section
over the space of gauge orbits and we would be done. If 
Berry's connection vanished, Berry's curvature, an 
exact two-form,
would also vanish. Berry's curvature is 
however always gauge
invariant, it is a property of the collection of the vector
spaces themselves, independent of basis choices. Therefore,
Berry's curvature two-form 
can be taken over the space of gauge
orbits, no matter what section we started with. 
Generically,
it does not vanish. Berry's 
curvature is entirely determined
by $H_W$. 
Therefore, the first question to ask is whether one can
deform $H_W$ in such a way that Berry's curvature vanishes.
\footnote{The two form is unchanged under $H_W\to g(H_W )$
where the real continuous function $g$ satisfies 
$\varepsilon (g(x)) = \varepsilon (x)$. 
It might be
convenient to search for the new $H_W$ 
in the form $g(H_W) + \delta$,
where $\delta$ is a small deformation. }
Here the latitude we allowed ourselves 
in the choice of $H_W$
gets exploited. 
By checking some examples it was 
found that there are cases where 
no small deformation of $H_W$ can make Berry's
curvature vanish. The obstruction shows up as follows:
One picks a special 
two dimensional, non-contractible compact
submanifold in the space of gauge orbits and shows that
the integral of Berry's two form over it gives a nonzero
integer. These integers turn out to
govern the continuum
anomalies of the chiral 
fermions one tries to discretize. 
This might be the deepest
geometrical feature of the overlap construction. 
For the first
time in lattice field theory 
one sees a purely geometrical role played by anomalies,
directly on finite lattices !

On the basis of these findings I conjecture that if 
continuum
anomalies cancel it is possible to deform $H_W$ in 
such a way
that the associated Berry curvature vanishes. I 
also conjecture that if Berry's 
curvature vanishes there is
a smooth section through the determinant bundle over the
space of gauge orbits. 
Assuming that these conjectures
are correct, there still remains 
the question whether an
explicit expression for $H_W$ and for the smooth sections
can be written down. It seems that even if constructive
proofs are found the explicit formulae would be too
complicated to be usable in numerical work. Nevertheless,
I don't think numerical work is 
condemned to be impracticable,
so long as the main goal is limited
to approaching the correct continuum limit. 

I believe that, as far as Physics goes, 
an explicit construction is not
really necessary. There are good reasons to believe that
even if one integrated a weight that was not exactly gauge 
invariant over all connections the result would still
have the correct continuum limit. The integration along
the orbit can be viewed as an averaging over the compact
group of gauge transformations. The integration over 
connections can be split into 
two stages, where in the first one 
averages over gauge orbits, 
producing a gauge invariant weight
to be further integrated over orbits. Even if one used an
$H_W$ for which Berry's 
curvature did not vanish and a section
in the bundle over connections that was not gauge invariant
the net result would have the same continuum limit as one
would get with a carefully chosen $H_W$ for which Berry's
curvature does vanish. It suffices that the actual $H_W$
is sufficiently close to the finely adjusted one. The
basis of this belief is a certain generalized universality
first proposed by F{\" o}rster, Nielsen and Ninomiya on
the basis of investigating some toy models. Indeed,
it would be unnatural if the correct continuum limit
depended on whether one picks a complicated
exact $H_W$ or just an approximation. 

An interesting degenerate case holds for $SU(2)$, where
a related issue, first pointed out by Witten, 
arises for Weyl fermions in pseudoreal representations.
All the spaces we dealt with can be 
taken over the reals, and
one ends up with a $Z(2)$ determinant 
bundle rather than a $U(1)$
one. In this case one has no Berry curvature, just
Berry holonomies. Explicit examples with nontrivial
irremovable Berry holonomy were constructed when anomalies
are known to occur in the continuum. The nontrivial
holonomy was 
found by computer, using a well known relation 
between degeneracies in $H_W$ and Berry holonomy. 

\section {Numerical Methods}

For vectorial theories  
any reasonable $H_W$ would work. But,
one needs to make it as sparse as possible if one
wants a practical procedure. This is of direct interest
in QCD, a vectorial gauge theory describing strong
interactions. 

In Nature, the ``strong interactions'' become strong only
at relatively low energies while all unknown physics
resides at high energies. One might think that one
should not be in the way of the other. 
But, because of their strength, the strong interactions
mask numerically almost 
any other property of strongly interacting particles.
It is important to be able to calculate these ``masking
factors'' in order to separate out 
potentially interesting new Physics
from the measured quantities. There is only
one method based on first principles to do the needed 
calculations. This is the specialty of the 
subfield of numerical lattice QCD. 
As I already mentioned,
in Nature, there are almost exact global chiral symmetries
and one needs to reproduce them one way or another on the
lattice. Recent developments indicate that this might
be now doable much more elegantly than before. 

In practice, one uses almost exclusively Krylov space
methods, since the matrices $H_W$ are of the order
of $10^6 \times 10^6$ and more. Only their sparseness
saves the day and the numerical 
methods one can at all consider
must use only matrix acting on vector operations. 

At the moment the methods of choice use some approximant
for the sign function $\varepsilon (x)$. 
One requires that numerical
work proceed with relative efficiency, 
while, at the same time, the 
approximation be of reasonable quality. 
Almost all methods essentially use a 
rational polynomial approximant. 
It is unclear at present against 
what criteria should one optimize
the approximant $\varepsilon_n (x)$ to the sign function. 
A nice coincidence is
that in the area of control theory, applied
mathematicians have been faced with the need to deal with
the $\varepsilon (A)$ for matrices 
A (the Roberts sign function). 
What is special to our case is that 
our A is hermitian,
huge, but sparse. A possible representation is
$\varepsilon(x)=\lim_{n \to \infty} 
\varepsilon_n (x)$ with
\be
\varepsilon_n (x)={{(1+x)^{2n} - (1-x)^{2n}}\over
{(1+x)^{2n} + (1-x)^{2n}}}=
{x\over n} \sum_{s=1}^n {1\over
{\cos^2 [(s-{1\over 2}) {\pi\over {2n}}]x^2
+\sin^2 [(s-{1\over 2}){\pi\over {2n}}]}}\ee

The action of the sum over poles on a given vector
can be computed by a slight 
generalization of the Conjugate
Gradient algorithm which uses the same Krylov space
for all $n$ inversions, and costs numerically not much
more than a single inversion. This is enough to
study various properties of $D_o$, and also of its inverse.
Since $D_o$ needs to be inverted, we end up with a two
stage nested invertor. I do not know whether 
the errors in exact arithmetic
and, moreover, the errors in real 
arithmetic, of such a nested
algorithm have ever been studied in the applied 
mathematics literature. 

To include the determinant in simulations seems at present
too expensive numerically if one must use the nested
algorithm. But, there may be a way out, using a continued
fraction representation of the 
approximants $\varepsilon_n (x)$. 
We want the determinant
of $2H_o=\epsilon^\prime +\epsilon_n (H_W)$ 
in the limit of
large $n$. 
We add some extra fields, 
and introduce a new 
bilinear ``action'' $S_0$:
\be
S_0 =\bar\psi\gamma_5\psi +\bar\psi\bar 
A_1\phi_1 -\bar\phi_1 A_1 \psi +
\bar\phi_1 B_1 \phi_1 + \dots + \bar\phi_{m-1} 
\bar A_m \phi_{m} -
\bar\phi_m A_m \phi_{m-1} +
\bar\phi_m B_m \phi_m \ee
Fields with bars are rows, 
without are columns, and in-between
we have square matrices. 
We write $S_0 =\bar\chi Q\chi$ with
$\chi = \pmatrix {&\psi\cr&\phi_1\cr 
&\vdots\cr &\phi_m\cr},~\bar\chi = (\bar\psi,
\bar\phi_1,\dots,\bar\phi_m)$. 

Simple manipulations show that if the $A_i ,
\bar A_i , B_i$ are commuting matrices 
($\det B_i \ne 0$), we have 
$\det Q =\prod_{i=1}^m (\det B_i ) 
\det (\epsilon^\prime +R)$
where,
\be
R={ {\displaystyle A_1\bar A_1}\over \displaystyle B_1 +
                 {\strut {\displaystyle A_2\bar A_2 } 
\over\displaystyle B_2 +
                   {\strut {\displaystyle A_3\bar A_3} 
\over\displaystyle B_3 +
                 \dots    {\strut \displaystyle  \ddots 
\over \displaystyle
B_{m-1} +
                     {\strut {\displaystyle A_m \bar A_m } 
\over
    \displaystyle B_m }}}}}
\ee
For the 
choice of $\varepsilon_n (x)$ in equation (1) 
we have a representation that goes back to Euler,   

\be
\varepsilon_n (x) =
{ {\displaystyle 2nx }\over \displaystyle 1 +
                 {\strut {\displaystyle (4n^2-1)x^2 } 
\over\displaystyle 
                                      3+
                   {\strut {\displaystyle (4n^2-4)x^2 } 
\over\displaystyle 
                                      5 + \dots
                   {\strut \displaystyle   \ddots \over 
{\displaystyle 4n-3}+
                     {\strut {\displaystyle 
[4n^2 - (2n-1)^2]x^2} \over 
{\displaystyle 4n-1} }}}}}\ee

Therefore, one can implement the 
rational approximation
by extending the matrix size. One can use 
other continued
fraction representations and other
approximants; for example new approximants
can be generated by exploiting $\varepsilon(x)
=\varepsilon (\lambda x)$ for $\lambda >0$
or $\varepsilon(x)
=\varepsilon ({1\over x})$ for
$x\ne 0$.
In some of the new continued fractions
the $B_i$ matrices may
become polynomial in $H_W$. In this case one 
needs to compensate
for the prefactor $\prod \det (B_i) $ which 
now carries some
dependence on gauge orbits. 
The compensation is easily
done stochastically, using ``pseudo-fermions''. 
Pseudo-fermions
are numerical integration 
variables that carry indices of the same kind as
carried by ordinary Grassmannian fermions. 

More developments in the area of 
algorithms as applied
to the overlap 
are expected as the subject is relatively young.

\section{Summary}

The subject of chirality on the 
lattice seems to require a 
wide range of tools from Mathematics. At the one
end one has algebraic topology, and at the other
numerical analysis. On the way one goes 
through principal
bundles, index theorems, 
various line bundles, to approximation 
theory, control theory, orthogonal polynomials, continued
fractions, stochastic processes, 
and numerical linear algebra.  Lattice
field theory could well use the help of professional
mathematicians. I hope more will get interested and
involved.   

\section{Guide to literature}

Rather than including detailed references 
I choose to present
a shorter list of mainly review papers. 
The list is subjective. 
My recommendation for the prime source to 
learn about chiral
gauge theories in the continuum are the lectures of L.
Alvarez-Gaum{\'e} \cite{lag}. 
An important second source I suggest
is the review of R. D. Ball \cite{rdb}. 
For Berry phase topics
I recommend the insightfully annotated collection of papers
edited by A. Shapere and F. Wilczek \cite{asfw}. 
A recent brief
summary of overlap work from the Physics 
perspective can be
found in my contribution to ICHEP98 \cite{ichep}. 
A more extended up to date
review about overlap work does not exist at the moment; the
closest is an older summary paper 
I wrote with R. Narayanan \cite{ovlap}.

\section*{Acknowledgment} 
This research was 
supported in part by the DOE under grant \#
DE-FG05-96ER40559.

\end{document}